# Room-temperature giant Stark effect of single photon emitter in van der Waals material

Yang Xia[1*], Quanwei Li[1*], Jeongmin Kim[1], Wei Bao[1], Cheng Gong[1], Sui Yang[1], Yuan Wang[1] and Xiang Zhang[1,2†]

[1] NSF Nanoscale Science and Engineering Center (NSEC), 3112 Etcheverry Hall, University of California, Berkeley, California 94720, USA.

[2] Materials Sciences Division, Lawrence Berkeley National Laboratory, 1 Cyclotron Road, Berkeley, California 94720, USA.

[*] These authors contributed equally to this work.

[†] e-mail: xiang@berkeley.edu

**Single photon emitters (SPEs) are critical building blocks needed for quantum science and technology[1]. For practical applications, large-scale room-temperature solid-state platforms are required[2,3]. Color centers in layered hexagonal boron nitride (*h*BN) have recently been found to be ultra-bright and stable SPEs at room temperature[4]. Yet, to scale up solid-state quantum information processing, large tuning range of single photon energy is demanded for wavelength division multiplexing quantum key distribution, where indistinguishability is not required, and for indistinguishable single-photon production from multi-emitters. Stark effect can tune the single photon energy by an electric field, which however, has been achieved only at cryogenic temperature so far[5–8]. Here we report the first room-temperature Stark effect of SPEs by exploiting *h*BN color centers. Surprisingly, we observe a giant Stark shift of single photon more than 30 meV, about one order of magnitude greater than previously reported in color center emitters[7–11]. Moreover, for the first time, the orientation of the electric permanent dipole moment in the solid-state SPE is**

**determined via angle-resolved Stark effect, revealing the intrinsic broken symmetries at such a color center. The remarkable Stark shift discovered here and the significant advance in understanding its atomic structure pave a way towards the scalable solid-state on-chip quantum communication and computation at room temperature.**

Van der Waals (vdW) materials, ranging from semi-metallic graphene[12] and semiconducting transition metal dichalcogenides[13] to insulating hexagonal boron nitride (*h*BN), have enabled remarkable scientific and technological breakthroughs over the last two decades thanks to their exceptional electronic and optical properties. Both the single materials and the heterostructures have been exploited to demonstrate appealing device applications[14], such as light emitting diodes[15], lasers[16] and optical modulators[17]. While most cases deal with classical information, only a few studies have been reported in the quantum regime at liquid helium temperature[18–23]. Recently, color centers in *h*BN have emerged as superb room-temperature solid-state single photon emitters (SPEs)[4], which opens up the possibilities of utilizing vdW materials as a platform for room-temperature solid-state quantum information systems. They are capable to work at room temperature, and among the brightest SPEs due to its high internal quantum efficiency. Moreover, high-efficiency photon extraction can be greatly facilitated by their intrinsic layered material structures[4]. Consequently, millions of linearly polarized photons per second can be easily detected without additional photon extraction structures. Furthermore, their facile integration with photonic and electrical components is highly preferred for integrated on-chip quantum information systems[13,14,24].

One major challenge for all solid-state single photon emitters is the random variation of emission energy caused by the inhomogeneity in local environment. Such variation breaks the indistinguishability of single photons from multiple emitters, which is critically required for large scale quantum computation, such as universal linear optics[25] and boson sampling[26]. The randomness also prevents scaling up the room temperature quantum communication systems from using wavelength-division multiplexing (WDM), due to the stringent requirement on the precision of photon energy placed by the narrow-band optics, although photon indistinguishability is not required for such applications[3]. Stark effect, which describes the shift of spectra lines by an external electric field (Fig 1c), can precisely control SPE photon energy and be facilely integrated into quantum systems[6], advantageous over other tuning methods such as temperature[27], strain[28] and magnetic field[29]. The Stark effect (Fig 1c) has been used to tune the emission energy of quantum dots[6], single photon emitters in layered $WSe_2$[5], atomic emitters such as NV centers and SiV centers in diamond[7–9,11] and organic dye molecules[10], all at liquid helium temperature, because either the emitters only produce single photons at low temperature, or the effect was too weak to observe at room temperature.

In this paper, we report the first room-temperature Stark effect of an SPE, discovered in an $h$BN color center. Surprisingly, we observe a giant tuning range around 31 meV at 300 K which is one order greater than previously reported in color center SPEs[30]. Moreover, we develop a characterization method to resolve the angular dependence of the Stark effect to determine the underlying symmetry of the color center. With this method, we directly observe, for the first time, a dipolar pattern of the Stark shift that is well aligned with the optical polarization. This dipolar pattern unambiguously reveals an electric permanent dipole moment which proves the breaking of inversion and rotation symmetries at the $h$BN SPE. The discovered remarkably giant room-temperature Stark effect and the significant advance in understanding its atomic structure could enable new possibilities of quantum information technologies, such as WDM and indistinguishable single photon sources, at room temperature.

To achieve large Stark shift of our $h$BN SPEs and to fully characterize its dependence on the amplitude and orientation of the local electric fields, we design the nano-scale four-electrode device (Fig. 1). SPEs in multi-layer $h$BN nano-flakes are chosen due to their much better optical performance compared to those in monolayers[4]. Multiple electrodes are carefully designed to surround the SPEs such that we can control not only the amplitude but also the direction of the electric field. Taking the advantage of the nanofabrication, we are able to develop a down-scaled four-electrode device with gaps of 200 nm and 400 nm between the adjacent and diagonal electrodes, respectively, to achieve a large electric field of ~ 0.1 V/nm, orders of magnitudes higher than previous reports[8,30]. Fig. 1a, b shows microscope images of the fabricated four-electrode device. We locate the SPE on the $h$BN flake by a localization analysis of its photoluminescence (PL) profile with respect to the electrodes (details in Extended Data Figure 1).

The high-quality single-photon emission from the hBN color center is verified by PL spectroscopy at room temperature before applying external electric fields (Fig 2a). The majority of its PL emission is attributed to the zero-phonon line (ZPL) at 2.088 eV. The narrow full width at half maximum (~ 7 meV) provides evidence as a high-quality emitter. Two small phonon sidebands (PSBs) are observed at 1.921 eV (PSB1) and 1.753 eV (PSB2), with the frequency difference of ~1370 cm$^{-1}$ corresponding to the E$_{2g}$ phonon of $h$BN[31]. The well resolved doublet at PSB1 is a typical feature for $h$BN nanoflake[15–17]. A few tiny peaks are visible that might result from the PL emission of other color centers in the collected region. The emission of the $h$BN SPE is linearly polarized (blue circles and curve in Fig. 3b), which fits well to a cosine-squared function with a visibility of 0.72. Many SPEs are characterized under the identical pump laser polarization and the detected photons are linearly polarized in various directions, thus the polarization observed here is specific to the SPE and not due to optical excitation. We measure the second-order coherence function (g$^{(2)}$) after device fabrication using a Hanbury Brown and Twiss (HBT) setup, from which single photon emission is confirmed by a raw anti-bunching dip of g$^{(2)}$(0) = 0.45 (Fig. 2a inset). By fitting the g$^{(2)}$ data to a single exponential decay function, we estimate the lifetime of our single photon emitter to be 4.2 ns.

After characterizing and confirming the SPE optical properties, we apply voltages within ±100 V between the electrodes A and B (Fig. 2b inset) to study the Stark effect. PL emission spectra are collected at each voltage. Fig. 2b plots the ZPL PL intensity map as a function of photon energy and applied voltage. A huge Stark shift of 31 meV is clearly observed, about one order of magnitude greater than the values previously achieved in color center SPEs[7,8,30] and 4-fold larger than its room-temperature linewidth (~ 7meV). To further analyze the effect, we extract the ZPL peak position as a function of applied voltage in Fig 2c. The Stark shift is approximately linear to the applied voltage with a tuning efficiency of 137 μeV/V and reverses sign at opposite electric field, which suggests a non-zero electric permanent dipole moment at the color center. The presence of degenerate states is ruled out, because there is neither ZPL splitting nor multi-peaks with applied fields (even at different excitation polarizations)[8]. The deviation from linearity is possibly due to the light induced ionization in nearby non-emitting defects that can modify the local electric field[8]. A gradual decrease in its intensity is noted when voltage changes from -100 V to 100 V (Extended Data Fig. 2a), which is possibly due to the change of coupling to dark state as reported previously in diamond NV center[7]. The repeatability and stability of such room-temperature giant Stark effect is further characterized in multiple emitters as shown in Extended Data Figures 3-7.

Stark shift depends not only on the magnitude of the applied field, but also its orientation[32]. The underlying symmetry of the atomic structure of the color center thus can be probed by the angular dependence of the Stark effect. We further develop a Rotating Field Method that fully control both the magnitude and orientation of the total electric field at the location of SPE by assigning the voltages of multiple electrodes (Methods). With a fixed local electric field magnitude of 0.08 V/nm, the ZPL is 1.13 meV red (1.32 meV blue) shifted when the applied field pointing to 140° (320°), while the shift with the electric field along 230° is negligible. The Stark shifts $h\Delta v$ as a function of the angle $\theta$ of the local field **F** is well-fitted with electric permanent dipole moment model (Fig 3b) based on perturbation theory to the first order:

$$h\Delta v = -\Delta\boldsymbol{\mu} \cdot \mathbf{F} = -|\Delta\boldsymbol{\mu}||\mathbf{F}|\cos(\phi-\theta) \quad (1)$$

where $\Delta\boldsymbol{\mu}$ and $\phi$ denote the dipole moment responsible for the Stark effect and its orientation angle, respectively. Such result further justifies that the Stark effect is dominated by an electric permanent dipolar term. From the fitting we estimate the magnitude of the dipole is $|\Delta\boldsymbol{\mu}| = 0.65 \pm 0.04$ D where 1 D = 3.33e-30 Cm, which is on the same order as the NV center in diamond[7]. The discovered electric permanent dipole moment corresponds to the asymmetric charge distribution at the $h$BN SPE, which will facilitate the future study of the atomic structure and electronic levels of the color center. In contrast, the linear polarization of emitted photons characterized in earlier reports is determined only by the optical transition dipole and features cosine-squared angular dependence, which omits the information for inversion symmetry of atomic structure (see Methods). We clarify such distinctive difference in Fig. 3b (blue circles). It should be noted that the direction for maximum Stark shift is coincident with that of the emission polarization (Fig. 3b Inset), which corresponds to the intersection of the mirror symmetry plane of the color center and the $h$BN atomic layer plane.

We emphasize that the key to observe the room-temperature giant Stark shift is a combined effort of several crucial factors. First, the large band gap of $h$BN crystal and low phonon scattering make a superb room-temperate SPE. Second, the layered structure of $h$BN likely leads to an in-plane dipole moment[4], such that an applied in-plane electric field can be well aligned with the dipole orientation. Third, closely spaced electrodes (within a few hundreds of nanometers) enabled by nanofabrication can easily be integrated with the nano-scale $h$BN flake such that large voltages (even above 100V) can be applied for local electric fields (on the order of 0.1V/nm) around one order of magnitude greater than previous works in diamond color centers[30].

To conclude, for the first time, we report the room-temperature Stark effect of an SPE, with a giant shift of 31 meV in *h*BN color center. The demonstrated large and reliable energy tuning and modulation is important for room-temperature quantum applications such as wavelength-division-multiplexing for practical quantum communication. Moreover, for the first time we uncover the intrinsic broken symmetries of color centers in *h*BN by introducing the Rotating Field Method to enable angular mapping of the Stark effect. Our results provide fundamental knowledge for the understanding and applications of color centers in vdW materials and open a new route towards scalable solid-state quantum information systems at room temperature.

**Methods**

**Device fabrication.** The $h$BN nanoflakes are purchased from Graphene Supermarket, in the form of a liquid suspension and drop-cast on silicon substrate with ~ 280 nm thermal oxide on top. The samples are then annealed at 1000 C for 30 min in an Ar/H$_2$ environment followed by slow cooling down. Individual single photon emitters are found by fluorescence microscopy and characterized by PL emission spectroscopy, polarization analysis and g$^{(2)}$ measurement. Electron beam lithography is used to define the electrode pattern around the located single photon emitters. The electrodes are made of 5 nm Ti and 100 nm Au deposited by electron beam evaporation.

**Rotating Field Method.** When we apply an electric field via multiple electrodes, the total external field can be considered as linear combination of the fields generated by individual electrodes. In our experiment, two voltage signals are applied to electrodes A and C (Fig. 3b inset), while the other two electrodes and substrate are grounded. From linear combination, we have the equation below:

$$\begin{bmatrix} E_x \\ E_y \end{bmatrix} = \mathbf{K} \begin{bmatrix} V_A \\ V_C \end{bmatrix},$$

where $E_x$ and $E_y$ are the x and y components of external electric field at the single photon emitter location, $\mathbf{K}$ is a 2-by-2 matrix, and $V_A$ and $V_C$ are the voltages applied to electrodes A and C respectively. Matrix $\mathbf{K}$ is obtained from 3D FEM simulation (COMSOL). The simulated geometry is extracted from the real device. In order to obtain the first column of $\mathbf{K}$, we assign $V_A$ = 1 V and ground electrodes B, C, D. The obtained $E_x$ and $E_y$ form first column of $\mathbf{K}$. And the second column can be calculated similarly.

In order to generate a local field with specific amplitude $F_0$ and direction $\theta$, we consider the following equation:

$$\begin{bmatrix} E_x \\ E_y \end{bmatrix} = \frac{F_0}{L} \begin{bmatrix} \cos(\theta) \\ \sin(\theta) \end{bmatrix}.$$

Here we follow previous works and use Lorentz approximation to calculate the local field from the external field. $L = (\varepsilon_r+2)/3$ is the Lorentz factor. The relative permittivity $\varepsilon_r$ is taken from a reference[31]. Combining the two equations above gives

$$\begin{bmatrix} V_A \\ V_C \end{bmatrix} = \frac{F_0}{L} \mathbf{K}^{-1} \begin{bmatrix} \cos(\theta) \\ \sin(\theta) \end{bmatrix}.$$

**The structure information from the discovered electric permanent dipole moments.** The electric permanent dipole moments correspond to the charge distributions of the electronic states of the SPEs. It can be calculated for the ground and excited states of an SPE as

$$\boldsymbol{\mu}_{g,e} = \int \psi_{g,e}^* e\mathbf{r} \psi_{g,e} dr^3$$

where $\psi$ is the wave function of electronic states, and the subscriptions g and e correspond to ground and excited states, respectively. Such dipole moments contribute to Stark shift through $\boldsymbol{\Delta\mu} = \boldsymbol{\mu_e}-\boldsymbol{\mu_g}$ (Eq (1)). A non-vanishing $\boldsymbol{\Delta\mu}$ indicates non-zero $\boldsymbol{\mu_e}$ and/or $\boldsymbol{\mu_g}$, which must result from non-inversion symmetric probability densities of electrons $|\psi_e|^2$ and/or $|\psi_g|^2$ as well as atomic structure.

On the contrary, the optical polarization is determined by optical transition dipole moment $\boldsymbol{\mu}_{e \to g} = \int \psi_g^* e\mathbf{r} \psi_e dr^3$, which emits an optical wave with electric field parallel to the dipole, along the directions normal to it. After polarizer, the detected optical intensity has a squared-cosine dependence $I(\theta) = I_0\cos^2(\theta)$ on polarization angle $\theta$, which returns itself after $\theta \to \theta+180°$. As such measurement is always inversion symmetric, it cannot tell whether inversion symmetry breaks or not at the emitter.

**Data availability.** The data that support the findings of this study are available from the corresponding author upon reasonable request.

**References**

1. Charles Santori, David Fattal & Yoshihisa Yamamoto. *Single-photon devices and applications*. (Wiley-VCH, 2010).

2. O'Brien, J. L., Furusawa, A. & Vučković, J. Photonic quantum technologies. *Nature Photonics* **3,** 687–695 (2009).

3. Aharonovich, I., Englund, D. & Toth, M. Solid-state single-photon emitters. *Nat Photon* **10,** 631–641 (2016).

4. Tran, T. T., Bray, K., Ford, M. J., Toth, M. & Aharonovich, I. Quantum emission from hexagonal boron nitride monolayers. *Nat Nano* **11,** 37–41 (2016).

5. Chakraborty, C. *et al.* Quantum-Confined Stark Effect of Individual Defects in a van der Waals Heterostructure. *Nano Lett.* **17,** 2253–2258 (2017).

6. Dots, N. Q. Quantum-Confined Stark Effect in Single CdSe. *Science* **278,** 2114–2114 (1997).

7. Tamarat, P. *et al.* Stark Shift Control of Single Optical Centers in Diamond. *Phys. Rev. Lett.* **97,** 083002 (2006).

8. Bassett, L. C., Heremans, F. J., Yale, C. G., Buckley, B. B. & Awschalom, D. D. Electrical Tuning of Single Nitrogen-Vacancy Center Optical Transitions Enhanced by Photoinduced Fields. *Phys. Rev. Lett.* **107,** 266403 (2011).

9. Bernien, H. *et al.* Two-Photon Quantum Interference from Separate Nitrogen Vacancy Centers in Diamond. *Physical Review Letters* **108,** (2012).

10. Lettow, R. *et al.* Quantum Interference of Tunably Indistinguishable Photons from Remote

Organic Molecules. *Phys. Rev. Lett.* **104,** 123605 (2010).

11. Sipahigil, A. *et al.* Quantum Interference of Single Photons from Remote Nitrogen-Vacancy Centers in Diamond. *Physical Review Letters* **108,** (2012).

12. Geim, A. K. & Novoselov, K. S. The rise of graphene. *Nature Materials* **6,** nmat1849 (2007).

13. Xia, F., Wang, H., Xiao, D., Dubey, M. & Ramasubramaniam, A. Two-dimensional material nanophotonics. *Nat Photon* **8,** 899–907 (2014).

14. Geim, A. K. & Grigorieva, I. V. Van der Waals heterostructures. *Nature* **499,** 419–425 (2013).

15. Sundaram, R. S. *et al.* Electroluminescence in Single Layer MoS2. *Nano Lett.* **13,** 1416–1421 (2013).

16. Ye, Y. *et al.* Monolayer excitonic laser. *Nature Photonics* **9,** 733 (2015).

17. Liu, M. *et al.* A graphene-based broadband optical modulator. *Nature* **474,** 64 (2011).

18. Koperski, M. *et al.* Single photon emitters in exfoliated WSe2 structures. *Nat Nano* **10,** 503–506 (2015).

19. Tonndorf, P. *et al.* Single-photon emission from localized excitons in an atomically thin semiconductor. *Optica* **2,** 347 (2015).

20. Srivastava, A. *et al.* Optically active quantum dots in monolayer WSe2. *Nat Nano* **10,** 491–496 (2015).

21. He, Y.-M. *et al.* Single quantum emitters in monolayer semiconductors. *Nat Nano* **10,** 497–502 (2015).

22. Chakraborty, C., Kinnischtzke, L., Goodfellow, K. M., Beams, R. & Vamivakas, A. N.

Voltage-controlled quantum light from an atomically thin semiconductor. *Nat Nano* **10,** 507–511 (2015).

23. Palacios-Berraquero, C. *et al.* Atomically thin quantum light-emitting diodes. *Nature Communications* **7,** ncomms12978 (2016).

24. Pospischil, A. *et al.* CMOS-compatible graphene photodetector covering all optical communication bands. *Nature Photonics* **7,** nphoton.2013.240 (2013).

25. Knill, E., Laflamme, R. & Milburn, G. J. A scheme for efficient quantum computation with linear optics. *nature* **409,** 46–52 (2001).

26. Aaronson, S. & Arkhipov, A. The Computational Complexity of Linear Optics. in *Proceedings of the Forty-third Annual ACM Symposium on Theory of Computing* 333–342 (ACM, 2011). doi:10.1145/1993636.1993682

27. Reithmaier, J. P. *et al.* Strong coupling in a single quantum dot–semiconductor microcavity system. *Nature* **432,** 197–200 (2004).

28. Flagg, E. *et al.* Interference of Single Photons from Two Separate Semiconductor Quantum Dots. *Physical Review Letters* **104,** (2010).

29. Stevenson, R. M. *et al.* A semiconductor source of triggered entangled photon pairs. *Nature* **439,** 179–182 (2006).

30. Müller, T. *et al.* Wide-range electrical tunability of single-photon emission from chromium-based colour centres in diamond. *New J. Phys.* **13,** 075001 (2011).

31. Geick, R., Perry, C. H. & Rupprecht, G. Normal Modes in Hexagonal Boron Nitride. *Phys. Rev.*

**146,** 543–547 (1966).

32. Brunel, C., Tamarat, P., Lounis, B., Woehl, J. C. & Orrit, M. Stark Effect on Single Molecules of Dibenzanthanthrene in a Naphthalene Crystal and in a *n* -Hexadecane Shpol'skii Matrix. *The Journal of Physical Chemistry A* **103,** 2429–2434 (1999).

33. Akselrod, G. M. *et al.* Probing the mechanisms of large Purcell enhancement in plasmonic nanoantennas. *Nature Photonics* **8,** 835–840 (2014).

34. Grange, T. *et al.* Cavity-Funneled Generation of Indistinguishable Single Photons from Strongly Dissipative Quantum Emitters. *Physical Review Letters* **114,** (2015).

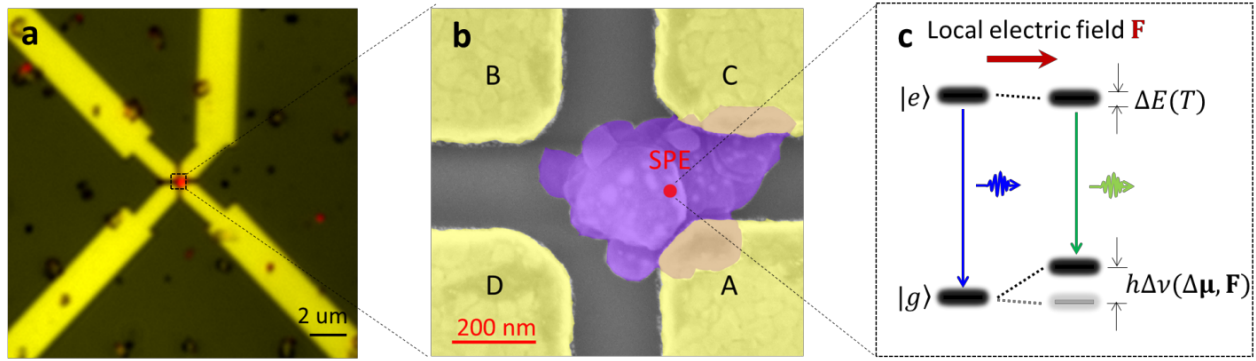

**Figure 1 | Device and physics of the Stark effect in *h*BN single photon emitter (SPE) at room temperature. a,** Two-channel optical image of the fabricated four-electrode device (Yellow: bright-field image of the gold electrodes; Red: photoluminescence (PL) image of the *h*BN SPE). **b,** Zoom-in pseudo-color SEM image of the same device. The *h*BN nano-flake (purple) hosts the SPE (red dot) whose position is found from a localization image analysis of (a) (details in Extended Data Fig. 1). A, B, C and D (yellow) denote the four gold electrodes where voltages are applied to generate external electric fields. **c,** Illustration of the Stark effect of the SPE (represented by a two-level system) with an optical transition from the excited state $|e\rangle$ to the ground state $|g\rangle$. The emitted photon energy is tuned via shifting the electronic levels by a local electric field **F**. At room temperature, the electronic levels and thus the emitted photon energy are broadened (characterized by $\Delta E(T)$) due to electron-phonon scattering, which sets the minimum Stark shift needed for practical use.

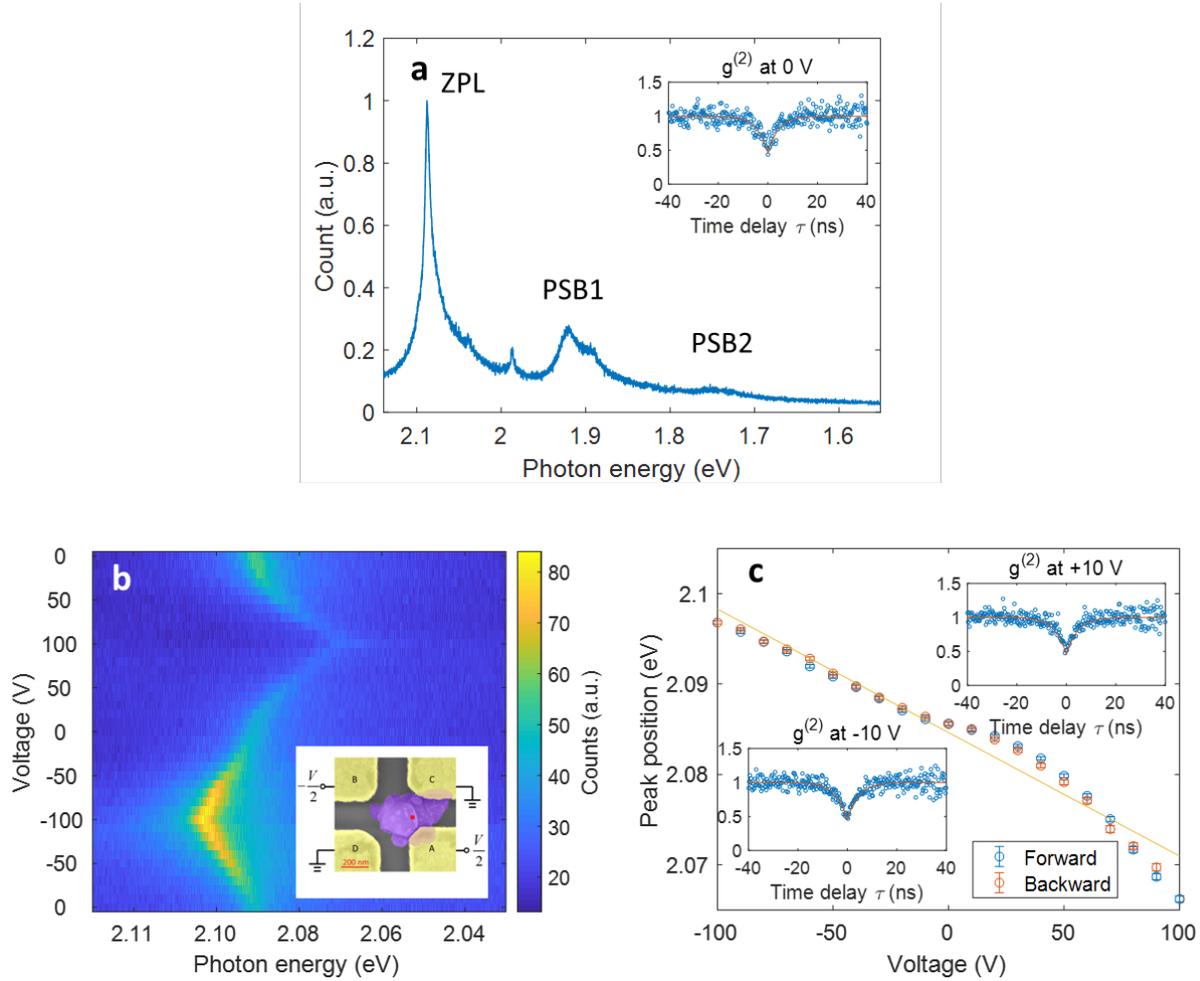

**Figure 2 | Observation of room-temperature giant Stark effect in hBN SPE. a,** Photoluminescence (PL) spectrum of the SPE at 300 K without applying electric field. It shows a dominating zero-phonon line (ZPL) at 2.088 eV with 7 meV full width at half maximum (FWHM) and two phonon sidebands (PSB1 and PSB2). The small peak at 1.988 eV stems from another emitter nearby. Inset: The measured (circles) and fitted (red curve) second order coherence function $g^{(2)}$ of the SPE PL after device fabrication. The $g^{(0)}(0)$ of 0.45 demonstrates the single-photon nature. The measurement data is well fitted by a single exponential decay with a lifetime of 4.2 ns. **b,** ZPL spectra of the *h*BN SPE as a function of voltage applied to electrodes A and B (inset) with equal magnitude at opposite signs. The achieved Stark shift is up to 31 meV, one order greater than previous reports in color center SPEs[7,8,30] and 4 times greater than its own room-temperature linewidth. **c,** Voltage controlled ZPL energy extracted from emission spectra

fitting from **b**. The blue and orange dots correspond to experiment data obtained during the forward and backward sweeping of voltages, respectively. Error bar, 95% confidence interval of the fitting. A tuning efficiency of 137 μeV/V is obtained by linear regression (yellow solid line). Insets show $g^{(2)}$ of the device measured at ±10 V, certifying that the single photon emission remains under external electric fields. The spectra and $g^{(2)}$ are measured under the excitations of continuous-wave 473 nm and 532 nm lasers, respectively. The acquisition time for $g^{(2)}$ is 10 s. The excitation intensity is 100 uW/μm² for all measurements.

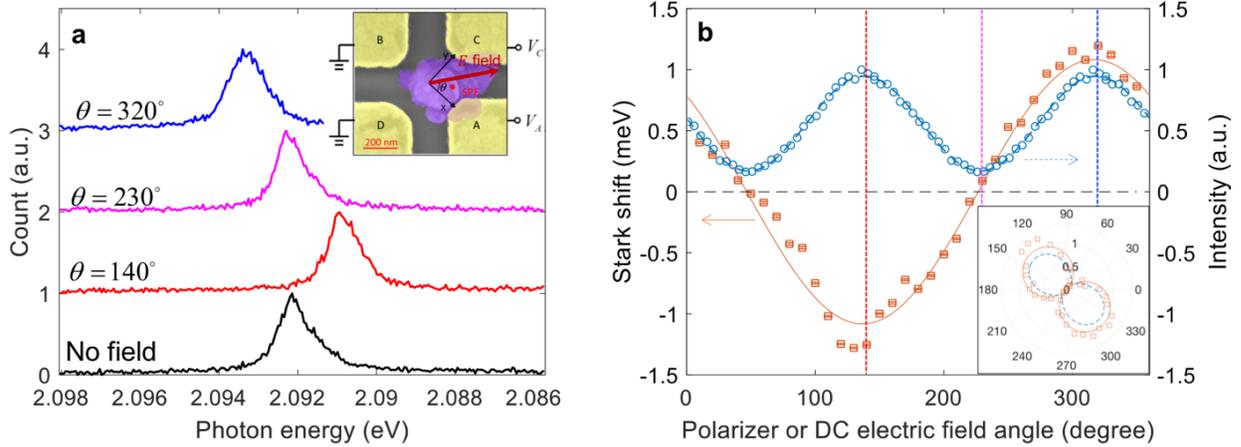

**Figure 3 | Angle-resolved Stark effect of *h*BN SPE and the discovered symmetry breaking. a,** ZPL spectra of *h*BN SPE recorded with electric fields applied in various orientations $\theta$ (defined in the inset) with a fixed magnitude (**F** = 0.08 V/nm). Such electric fields are generated by applying voltages to electrodes A and C (inset see Methods for details). The zero-field spectrum is plotted for comparison. When we apply electric field along 140° (320°) directions, 1.32 meV red (1.13 meV blue) shift of ZPL is observed. On the contrary, the electrical field along 230° does not cause noticeable change in ZPL spectrum. **b**, Angle-resolved Stark shift (orange color, left y axis) and optical polarization data (blue color, right y axis). The x axis corresponds to the orientation angle of the applied electric field and that of the polarizer in front of the photodetector, respectively. The orange squares (blue circles) are the measured Stark shift (ZPL intensity) of the single photons, and the solid orange line (dashed blue line) is the fitting curve according to the electric permanent dipole model in Eq. (1) in text (to the linearly polarized emission in cosine-squared function). The unveiled electric permanent dipole moment uncovers the broken inversion and rotation symmetries at the atomic color center. The electric permanent dipole moment aligns well with the emission polarization. The inset shows the same data in polar coordinates. Three vertical dashed lines in the main panel (blue, magenta and red) correspond to the three spectra in (a). The photon energies in (b) are obtained by fitting the ZPLs with the Lorentzian line shape. The error bars from fitting are smaller than 0.03 meV. The Stark shift is measured at 80 K to reduce ZPL fitting uncertainty at small shifts, taking advantage of

the narrow linewidth at low temperature. The optical polarization is measured at room-temperature.